\documentclass[report,floatfix,aps,nofootinbib]{revtex4}
\usepackage{epsfig}

\def\lp {\left( }
\def\rp {\right) }
\def\lb {\left[ }
\def\rb {\right] }

\def\nn {\nonumber}

\def\beq{\begin{equation}}
\def\eeq{\end{equation}}
\def\bea{\begin{eqnarray}}
\def\eea{\end{eqnarray}}
\def\ni{\noindent}

\def\m{\mu}

\begin{document}

\title{Quark confinement and curved spaces}

\author{C.C. Barros Jr.}

\affiliation{Instituto de F\'{\i}sica, Universidade de S\~{a}o Paulo,\\
C.P. 66318, 05315-970, S\~ao Paulo, SP, Brazil}

\date{\today}

\begin{abstract}
In this work it will be shown how quark confinement appears when wave equations
derived in curved spaces are considered. First, the equations and their
solutions  for Coulomb-like potentials will be presented, and
then, how this theory leads to quark 
confinement. A comparison between different models of confinement will be also 
made.

\end{abstract}

\maketitle

\vspace{5mm}

\section{introduction}

The introduction of quarks in the physical theory in  1964, by the  Gell-Mann 
\cite{gm1} and Zweig \cite{zw} hypothesis,  
has also introduced the annoying question of quark confinement. If by one hand 
this theory was able to organize and to explain the main proprieties of the 
hadrons with the scheme called the
eightfold way \cite{gm2}, on the other hand, 
the 
impossibility of observing
 free quarks could be considered as a major problem. Since then, many 
authors 
proposed models in order to describe the structure of the hadrons 
in terms of confined quarks. A successful way to implement these ideas
 is to consider nonrelativistic constituent quark  models, such as the 
 the nonrelativistic oscillator model,
proposed by Dalitz in 1967 \cite{dal} and by Faiman in 1968 \cite{fai}, where 
the 
Baryons are supposed to be systems composed of 
 three constituent quarks,  confined by an oscillator potential, with the 
states 
determined by the Hamiltonian
\beq
H_0=\sum_i {p_i^2\over 2m_i} + 
{K\over 2}\sum_{i>j}\lp \vec r_i -\vec r_j \rp^2  \ .
\label{hnr}
\eeq    

\ni
This model was improved by De R\'ujula \cite{a} with the addition of
a quark-quark 
spin-dependent potential,  by Karl and Isgur, with the introduction of a 
quantum chromodynamics inspired potential \cite{karl} and also by Murthy
 \cite{bhad} that considered a deformed oscillator.
Heavy $q\bar q$ systems are equally well described by nonrelativistic 
constituent quark models,
such as the Cornell model \cite{corn}, \cite{d}, 
that uses a linear plus Coulomb potential of the type
\beq
V(r)={a\over r}+ br  \  \  ,
\label{v1}
\eeq

\ni
where $a$ and $b$ are constants determined phenomenologically.
 Some models, as for example
 \cite{b}, \cite{c}, are based on other mechanisms, and generate
different potentials.

Despite  the success of the nonrelativistic models in describing the 
hadrons proprieties, theoretically, it is more reasonable  to think  
  quarks as relativistic particles. In 1968,
Bogolioubov \cite{bog} considered the Baryons as spherical cavities, and inside
of them the three
constituent quarks are
Dirac particles, submitted to a self-consistent mean 
field, that was represented by a  scalar potential
\bea
&&V(r)=0  \  \  {\rm for}  \  \  r\leq R   \nn \\
&&V(r)=V_0  \  \  {\rm for}  \  \  r > R   \  \  ,
\label{v2}
\eea

\ni
and quark confinement is achieved for $V_0\rightarrow \infty$.
Further development of these ideas leads to the MIT bag model \cite{mit} 
where the vacuum pressure was included. Other models based on the Dirac 
equation
\cite{x}-\cite{zz}, with  $r^n$ confining  potentials,
as in \cite{zz}
\beq
V(r)=V_0+\lambda r   \   ,
\label{v3}
\eeq

\ni
 may be 
found  in the literature, and they also show  good agreement with the 
experimental data.

In a recent paper \cite{cc}, a relativistic wave equation based on the general 
covariance principle has been derived. This theory has been constructed 
taking into account the effect of different kind of interactions 
(electromagnetic and strong) in the metric of the space-time. 
 In this work, the main objective is to 
investigate
 the quark confinement with this theory. As it will be seen in the next 
sections, very  interesting results can be obtained this way, and in many 
aspects 
these results are qualitatively different from the ones obtained in the 
previously cited models. 

This paper has the following structure:
In Sec. II a brief review of the theory and the solution of the equation for a
Coulomb-like potential are  shown,  in Sec. III, the quark confinement 
effect that comes from this theory  is presented, and in Sec. IV,  a 
comparison between the different
confinement mechanisms discussed in
this paper is  made and the conclusions are drawn.

\section{Quantum mechanics in curved space-time}

The  Einstein general theory of relativity is one great achievement
 in the understanding of Nature, and when applied to very large 
systems, such as planets or galaxies, gives very precise results.
Taking this fact into account, 
 fundamental questions may be asked, as for example
 why the general covariance principle
 does not apply to
very small systems, such as atoms or elementary
particles, and if the laws of physics depends on the size of the object.
In quantum 
systems,  the electromagnetic and strong interactions dominate and
the gravitational interaction is negligible, as the masses of the 
considered particles are very small. So, 
 the gravitational
potential may be turned off, and then, 
 the curvature of the space-time will be predominantly
 due to the other interactions 
(electromagnetic and strong).

With these 
aspects in mind, in \cite{cc} a theory was proposed, and
an equation similar to Dirac equation was
derived. In this section, a brief revision of this theory will be made, and 
some results, necessary to the development of this paper will be shown.

The simplest way to formulate this theory is to consider systems where
spherical symmetry exists. In this case,
the space-time is  described 
by the Schwarzschild metric \cite{lan},\cite{wein}, 
\beq
ds^2=\xi\ d\tau^2 - r^2(d\theta^2+ \sin^2 \theta\ d\phi^2) - \xi^{-1}dr^2  \ 
 \  ,
\label{I.1}
\eeq

\ni
where the factor
$\xi(r)=(1+V(r)/m_0c^2)^2$ is determined by the interaction potential 
$V(r)$, and is a function only of $r$. 

From the definition of the energy-momentum relations and the respective
 the quantum operators
(mathematical details may be found in \cite{cc})
in the given metric, the general relativistic  equation for spin-1/2 particles
\beq
{i\hbar \over \xi}{\partial\over\partial t}\Psi=\lp -i\hbar c\ 
\vec \alpha.\vec\nabla +\beta m_0c^2  \rp\Psi \  \  
\label{bar}
\eeq

\ni
has been deduced \cite{cc}, where $\Psi$ is a four-component spinor.
One must note that
despite of the fact that this theory is conceptually more complicated then 
the Dirac one,  the final equation is very similar to the
Dirac equation, what is surprisingly
in accord with his simplicity ideal.

 The spacial part of $\Psi$ may be 
written as
\beq
\psi=\pmatrix{F(r)\chi_k^\m\cr iG(r)\chi_{-k}^\m}  \  \  ,
\eeq

\ni
where $\chi_k^\m$ are the usual two-component spinors, and $k$ is related 
with the angular momentum by
\bea
&& k=l \ \  \  \  \   \  \   \  \  {\rm for} \  \  j=l-1/2 \  \  , \nn \\
&& k=-l-1  \  \  {\rm for} \  \  j=l+1/2  \  \   .
\eea

\ni
The radial part of 
 eq. (\ref{bar}) may be rewritten as a pair of coupled equations for the
and the $F$ and $G$ functions 
\bea
&& \sqrt{\xi} {dF\over dr}+(1+k){F\over r}= 
\lp{E\over \sqrt \xi} +m_0 \rp G  \nn \\  
&& \sqrt{\xi} {dG\over dr}+(1-k){G\over r}= 
-\lp{E\over \sqrt \xi} -m_0 \rp F  \  \  .
\label{dir}
\eea

Considering a coulomb-like potential $V(r)=-\alpha Z/ r$
the $\xi$ function becomes
\beq
\xi={\lp 1-{\alpha Z\over m_0c^2 \ r } \rp}^2    \  \  ,
\label{xep}
\eeq

\ni
and inserting it (\ref{xep}) in eq. (\ref{dir}) and
 making the substitution 
$\rho=\beta r$, the equations may be put in the form
\bea
&& \xi {dF\over d\rho}+ \sqrt{\xi}(1+k){F\over \rho}= 
\lp{E\over \beta} + \sqrt{\xi}{m_0\over \beta} \rp G  \nn \\  
&& \xi {dG\over d\rho}+\sqrt{\xi}(1-k){G\over \rho}= 
-\lp{E\over\beta} -\sqrt{\xi}{m_0\over \beta} \rp F  \  \  .
\label{dir2}
\eea

The  equations may be solved by the Frobenius method, expressing the
 $F$ and $G$ functions as power series
 of the form
\bea
&& F=\rho^s\sum_{n=0}^N a_n \rho^ne^{-\rho}   \  , \nn  \\
&& G=\rho^s\sum_{n=0}^N b_n \rho^ne^{-\rho}  \  . 
\label{frob}
\eea

\ni
Substituting this expressions in the equations
we find that $s=0$ and
 the relations between the 
coefficients are obtained
\bea
&& a_1=\lb{1+k+\alpha\beta\over\alpha\beta }  \rb a_0    \nn  \\
&& b_1=\lb{1-k+\alpha\beta\over\alpha\beta }  \rb b_0  \  \  ,
\label{c1}
\eea 
\bea
2\alpha^2\beta^2a_2 &-& \alpha\beta\lp 3+k+\alpha\beta \rp a_1 +  
\lp 1+k+2\alpha\beta  \rp a_0+    \alpha m_0b_0=0 
\nn \\
2\alpha^2\beta^2b_2 &-& \alpha\beta \lp 3-k+\alpha\beta \rp a_1 +
 \lp 1-k+2\alpha\beta  \rp b_0+\alpha m_0a_0=0 
\label{c2}
\eea

\ni
and
\bea
&&(n+3)\alpha^2\beta^2 a_{n+3} - 
\alpha\beta\lb 2n+ 5+k + \alpha\beta \rb a_{n+2} + 
\lb n+2+k+2 \alpha\beta \rb a_{n+1} - a_n+
 \alpha m_0b_{n+1}-{(E+m_0)\over\beta}b_n
=0  \nn \\
&& (n+3)\alpha^2\beta^2  b_{n+3} - 
\alpha\beta\lb 2n+5-k+\alpha\beta \rb b_{n+2} +
\lb n+2-k+2 \alpha\beta \rb b_{n+1} - b_n+ 
   \alpha m_0a_{n+1}+{(E-m_0)\over\beta}a_n   
=0 \nn \\
\label{c3}
\eea

\ni
with $a_0, a_1, a_2, b_0, b_1, b_2\neq 0$.
 From these relations, one obtains $\beta=\sqrt{m^2-E^2}$
that determines the factor $e^{-\sqrt{m^2-E_N^2}r}$ in the wave functions
(\ref{frob}) what determines the same behavior that was obtained with the
Dirac equation. The relation
\beq
\lb N+2\alpha\beta  \rb\beta -\alpha m^2=0  
\label{eqind}
\eeq 

\ni
is also obtained from eq. (\ref{c1})-(\ref{c3}), and 
 gives 
 the relation for the energy levels 
\beq
E_N= \pm m_ec^2 \sqrt{{1\over 2}- {N^2\over 8\alpha^2} \pm 
{N\over 4\alpha}\sqrt{{N^2\over 4\alpha^2}+2}}   \  ,
\label{sbr}
\eeq

\ni
where the physical values are the positive ones. 
 This spectrum may be compared with the
one obtained with the Dirac equation \cite{scf}, \cite{bj} 
(also deduced by Sommerfeld \cite{som})
\beq
E_N={m_ec^2\over \sqrt{1+\alpha^2/a^2}}  \  \  ,
\label{hsp}
\eeq

\ni
where $a=N-j+1/2+\sqrt{(j+1/2)^2-\alpha^2}$.

\ni
if one takes numerical results. For example, 
for the electron-proton interactions in
the deuterium atom,  the results may be obtained considering
  the fine-structure 
constant \cite{PDG}
$\alpha$=1/137.03599976,  $Z=1$ and the electron mass
 $m_0$=0.510998902 MeV/$c^2$  \cite{PDG}. 
The experimental ground state energy for the deuterium atom is 
$E_0=$-13.60214 eV \cite{nist}, calculating it with the Dirac spectrum
 (\ref{hsp}), one has
-13.60587 eV, and with eq.  (\ref{sbr}), -13.60298 eV. More numerical results
may be found in \cite{cc}.
Observing these results, one can see that the accord of both theories
with the deuterium experimental data is very good, but the results from
eq. (\ref{sbr}) are closer to the experimental data then the 
results from eq. (\ref{hsp}).
The results form the Dirac theory  (\ref{hsp}), one can see 
that the deviations from the data are of the order 
of 0.027\%. Considering the spectrum of eq. (\ref{sbr}), the 
 deviations are of the order
of 0.005\%, almost five times smaller, what is a significant 
improvement. The same pattern occurs when the other energy levels are 
compared \cite{cc}.

\section{Quark Confinement}

At this point it is instructive to investigate  the
solutions of the wave equations (\ref{dir})  for the Coulomb potential
near the classical horizon of events, at $r=r_0$ and apply these ideas
 to strongly interacting systems.
One must remark that the solution presented in Sec. II
 is not valid at the surface
$r=r_0$, and for this reason,  this case must be studied separately. 
The solution of the equation in the neighborhood of $r_0$, may be
given by  an expression similar to (\ref{frob}), but 
 replacing $\rho$ for $\rho-\alpha\beta /m_0$,
\bea
&& F=\rho^s\sum_{n=0}^N a_n \lp \rho - {\alpha\beta\over m_0} \rp^ne^{-\rho}
   \  , \nn  \\
&& G=\rho^s\sum_{n=0}^N b_n \lp \rho  - {\alpha\beta\over m_0} \rp^ne^{-\rho}
  \  . 
\label{frob2}
\eea

\ni
With this procedure, one finds that near the horizon of events
just one energy value is possible, $E=0$.
 The other conditions for the existence of a solution are  $k=0$ ($l=0$) and 
$s=-1-\alpha$, what means
an infinite discontinuity of the wave function at $r=r_0$. 
If this solution is discarded,
 the trivial solution $\psi(r_0)=0$ must be considered, what can be
interpreted as a
boundary condition at $r=r_0$. Consequently , this solution tells us that the
 space is
 divided in two parts
(fact that is true in both cases) inside and outside the horizon,
 which does not
communicate.  

So, if one considers hadrons composed of quarks, that generate a 
self-consistent field, described by a potential  (here, as an example
we considered a Coulomb potential), inside the horizon of events, 
the quarks may be described by  solutions of the type (\ref{frob})
with the energy levels given by the expression (\ref{sbr}). At $r\sim r_0$,
the  solution is (\ref{frob2}), that imposes the confinement of quarks 
inside this region. Classically thinking, the quarks are confined by
a trapping surface \cite{pen} that is generated by the potential.

Considering
the $\Upsilon$ meson for example, that  is a $b\bar b$ state, the 
theory may be applied just considering $b$ constituent quarks with mass
$m=$5.5 GeV, and a  Coulomb potential with
$\alpha=1.05$, what is a reasonable value for quark interactions.
From the expression (\ref{sbr}) one has 
$E_0=$ 9.47 GeV ($m_\Upsilon$=9.46 GeV), and $E_{max}=$11.0 GeV, that is the
mass of $\Upsilon$(11020). 
The quarks will be confined inside the region $r<r_0$=0.05 fm,
 that is a reasonable size for a core of a  meson. Some estimates 
of these quantities for other hadrons
may be found in Table I \cite{cc}.

\begin{table}[hbt]
\begin{center}
\caption{Values of the masses $M$ of the hadrons, composed of constituent
quarks of mass $m$ compared with the experimental ones 
\cite{PDG}.
The calculations are made with eq. (\ref{sbr}), obtained for
 Coulomb potentials with coupling $\alpha$.} 
\begin{tabular} {|c|c|c|c|c|c|}
\hline

		& $m$(GeV) 	& $\alpha$ 	
& $r_0$(fm) 	 & $M$(GeV) & $M_{\rm exp}$(GeV)
		\\ \hline
Nucleon($qqq$)& 0.38  & 1.60  & 0.83 & 0.938    & 0.938 (proton)\\ \hline
$J/\psi (c\bar c)$& 1.79  & 1.00  & 0.11 &3.10  & 3.10  	\\ \hline
$ \Upsilon (b\bar b)$& 	5.50  & 1.05  & 0.05 & 9.47  & 9.46 	\\ \hline
\end{tabular}
\end{center}
\end{table}

One must remark that in order to describe the spectra of the particles of 
Table I, other terms must be added in the potential. This fact
 (that is widely used \cite{a}-\cite{c}) may be understood
considering that
for short-range interactions many effects may occur,
  generating corrections to the potential.
Another factor that must be considered to improve the description is 
that
spherical symmetry is not the best one for $q\bar q$ mesons, and this fact 
must be corrected in future works.

In fact, the horizon of events is not an exclusive feature of the Coulomb 
potential, it may appear for any attractive potential, when the condition
\beq
\xi(r_0)=0   \   \   ,
\eeq

\ni
is satisfied, what occurs for $V(r_0)=-mc^2$. In these cases, the only energy 
value is
 $E=0$, and the solution will present the  discontinuity 
showed above,
\beq
\psi\propto {f(r)\over \lp r-r_0  \rp^\delta}
\eeq

\ni
with $\delta > 0$,
results that lead to quark confinement for general attractive potentials.

\section{Discussion of the results}

In this work it was shown how quark confinement appears when relativistic wave
equations in curved spaces are used.
 Now the obtained results will be 
compared with the results of the existing models.

As it was said in the introduction, many authors succeeded in explaining quark 
confinement with phenomenological potentials. In the nonrelativistic 
oscillator model \cite{fai}-\cite{karl}, the Hamiltonian (\ref{hnr}) leads to 
confining
oscillator-type wave functions  that contain a factor
$ e^{-\beta(r_1^2+r_2^2+r_3^2)}$,
where $\beta$ is a constant. In \cite{zz} with the 
potential (\ref{v3}), the 
approximate behavior of the wave function is 
$\psi \propto \Phi(r) e^{-\beta r^2}$ and
in  the Cornell model  \cite{corn}, where the potential of the type (\ref{v1})
is used, a similar behavior occurs.
In the Bogolioubov and 
in the MIT bag models that consider a potential of the type (\ref{v2}), the
 wave functions contain a factor
$ e^{-\beta \sqrt{m^2-E^2}(r-r_0)}$, and quark confinement
appears  
in the limit $V_0\rightarrow \infty$, where the wave function is constant for 
$r=R$ and 0 for $r>R$.

As it was seen in last section, quark confinement appears in a different
 way when the equations derived in curved spaces are used.
 Differently from the other models,
for an internal particle it is not possible to reach the surface $r=r_0$, as 
 $\psi(r_0)=0$. One must observe that even in the MIT bag 
model, with an infinite potential, a so strong condition is not reached.
The result of this condition is that
 the space-time is divided in two
 disconnected regions, inside and outside the surface. 
This fact is an intrinsic propriety of the space-time, due to attractive 
potentials, as for example the Coulomb potential, and there is no need of
introducing confining potentials to obtain this effect. In fact, 
potentials of the type (\ref{hnr})-(\ref{v3}), represent in an approximate 
way, in  plane space-time formulations, systems that are described 
in a natural way by 
curved spaces.

Another interesting aspect of the theory is that
classically it is expected the collapse at the origin, but here we are
dealing with quantum systems, and the uncertainty principle forbids this
collapse. The solution of the equation shows that the wave function
is 0 at the origin, confirming this statement.

The Dirac theory \cite{dir1}, \cite{dir2}
 introduced the special relativity in quantum mechanics,
so it is very reasonable to think that the next step is to formulate the 
quantum mechanics in terms of  the general relativity principles. 
The deuterium spectrum obtained in this way shows  that
the corrections of the energy levels, due to this  general formulation of
quantum mechanics (or general quantum mechanics) with the inclusion of the 
electric curvature of the space-time,
 provide a quite impressive 
agreement with the experimental data. Another
 strong evidence in 
 the  validation the theory is the quark confinement mechanism proposed in 
this paper, where the quarks are confined by a trapping surface, 
similar to the one defined by Penrose \cite{pen}. 
Conceptually, these are very important results, 
 as they show a successful way to join the quantum mechanics and 
the general relativity.

\begin{acknowledgments}
I wish to thank M. R. Robilotta. This work was supported by FAPESP.
\end{acknowledgments}




\end{document}